# Resolving the valence of Iron Oxides by Resonant photoemission spectroscopy

Hao Chen [1+, *], Yun Liu [2+], Hexin Zhang [2, 3], Shengdi Zhao [2, 4], Slavomir Nemsak [5], Haishan Liu [6], Miquel Salmeron[1, 7, *]

**Affiliations:**

1 Chemical Sciences Division, Lawrence Berkeley National Laboratory, Berkeley, CA 94720, USA

2 Key Laboratory of Urban Pollutant Conversion and Fujian Key Laboratory of Atmospheric Ozone Pollution Prevention, Institute of Urban Environment, Chinese Academy of Sciences, Xiamen 361021, China

3 Collaborative Innovation Center of Chemistry for Energy Materials, College of Chemistry and Chemical Engineering, Xiamen University, Xiamen 361005, China

4 University of Chinese Academy of Sciences, Beijing 100049, China

5 Advanced Light Source, Lawrence Berkeley National Laboratory, Berkeley, CA, 94720, USA

6 Department of Electrical and Computer Engineering, University of California at Riverside, Riverside, CA 92507 USA

7 Materials Sciences Division, Lawrence Berkeley National Laboratory, Berkeley, CA, 94720, USA

*Corresponding author. E-mail: haochen@lbl.gov (H.C.); msalmeron@lbl.gov (M.S.)

**Abstract**

Precisely determining the oxidation states of metal cations within variable-valence transition metal oxides remains a significant challenge, yet it is crucial for understanding and predicting the properties of these technologically important materials. Iron oxides, in particular, exhibit a remarkable diversity of electronic structures due to the variable valence states of iron ($Fe^{2+}$ and $Fe^{3+}$), however, quantitative analysis using conventional X-ray photoelectron spectroscopy (XPS) is challenging because of significant overlapping of the Fe2p spectra among different oxidation states. In this study, we leverage the intriguing case of Pt supported $FeO_2$ phase of monolayer thickness (ML) as a model system and employ Resonant Photoemission Spectroscopy (ResPES) to directly quantify the cation valence states and compositional ratios in this complex Fe oxide. Our results reveal that this ultrathin $FeO_2$ film (Pt-O-Fe-O), contrary to the +3 valence predicted by density functional theory (DFT), consists of an equal mixture of $Fe^{2+}$ and $Fe^{3+}$ cations, yielding an average valence of +2.5. Structurally, $FeO_2$ is likely derived from the $Fe_3O_4$ sublattice, featuring an octahedral Fe layer (50% $Fe^{3+}$ and 50% $Fe^{2+}$) bonded to upper and lower oxygen layers.

**1 Introduction**

Iron (Fe), the most common element on Earth by mass, is ubiquitous in nature and plays a crucial role in shaping our society[1]. The redox reactions between ferrous ($Fe^{2+}$) and ferric ($Fe^{3+}$) oxides are central to numerous processes, spanning biochemistry[2], geochemistry[3, 4], and heterogeneous catalysis[5, 6]. In the realm of heterogeneous catalysis, platinum (Pt) supported on iron oxides is frequently utilized[5-12]. Reduction treatments can partially reduce the iron oxide substrate, causing migration of ultrathin Fe oxide moieties and encapsulation of Pt nanoparticles[13-15], which can significantly boost catalytic reactivity compared to that of the noble metal alone[6], an effect known as strong metal support interactions[16].

Research on ultrathin Fe oxide films supported on Pt(111) substrates has been extensively conducted since the 1990s. Somorjai et al. initially identified the surface structure of $FeO_x$ layers on Pt(111), revealing that it adopts an FeO (111)-(1×1) stoichiometry at monolayer thickness [17].

Using scanning tunneling microscopy (STM), Salmeron et al. found that FeO/Pt(111) forms a moiré pattern as a result from the lattice mismatch between the FeO (111) overlayer and the underlying Pt(111) substrate[18, 19]. This structure transforms into two distinct phases at higher coverages: a (2×2) structure and a (√3×√3)-R30° structure, which were identified as an Fe-terminated surface of $Fe_3O_4$ (111) and alpha-$Fe_2O_3$[20], respectively. Additionally, partially reduced FeO films on Pt were observed by Besenbacher et al.[21-24], and other groups[6, 25-30], featuring ordered oxygen vacancy dislocations from coordinately unsaturated Fe atoms.

Recently, Freund et al. reported the formation of a new oxygen-rich phase of FeO, i.e., $FeO_2$, under severe oxidation conditions[31, 32]. This $FeO_2$ (O−Fe−O) phase was shown to catalyze CO oxidation on Pt(111) at 450 K, where the interfacial oxygen is reacted by CO and replenished by $O_2$ [31, 33, 34]. This new phase features an extra oxygen layer intercalated between Pt and FeO film of monolayer thickness (ML), transforming the FeO (Pt-Fe-O) to a $FeO_2$ (Pt-O-Fe-O) structure. Density functional theory (DFT) calculations suggest that Fe cations in the $FeO_2$ phase have a formal oxidation state of +3 because of electron back-donation to the Pt substrate[32, 35]. The $FeO_2$/Pt(111) structure was also confirmed by Bao et al. [26, 36, 37] and Wendt et al.[38]. However, apart from ambiguous results from Fe 2p X-ray photoelectron spectroscopy (XPS), due to the strong overlap of the peaks from various Fe oxidation states, an experimental measurement of the valence state of $FeO_2$ is still lacking[39].

In this study, we use resonant photoemission spectroscopy (ResPES) to identify the different iron oxides and present direct experimental evidence that the valence state of $FeO_2$ is +2.5, comprising equal amounts of $Fe^{2+}$ and $Fe^{3+}$ species, lower than the +3 valence state predicted by DFT. [32, 35] Our results also suggest that the $FeO_2$ phase originates from the $Fe_3O_4$ (111) sublattice, where the octahedral Fe layer is structurally bonded to the upper and bottom O layers. We highlight that ResPES enables precise determination of the ferrous and ferric composition ratio of Fe oxides directly under reaction conditions.

## 2 Methods

The experiments were carried out in two UHV systems. The first system is equipped with a low-temperature scanning tunneling microscope (LT-STM, Boson Instruments, Beijing) in a base pressure of less than $3\times10^{-10}$ mbar and sample preparation chambers. The STM images were obtained at 78 K using a Pt-Ir tip. The second system, equipped with a Scienta R4000 HIPP analyzer, is located at beamline 9.3.2 of the Advanced Light Source (ALS, Lawrence Berkeley National Laboratory) where X-ray photoelectron spectroscopy (XPS), Resonant photoemission spectroscopy (ResPES), and partial electron yield X-ray adsorption spectroscopy (PEY-XAS) measurements at 200 eV electron kinetic energy were performed, respectively. The Pt(111) single crystal was cleaned by cycles of $Ar^+$ ion sputtering (2 keV, 10 μA) and annealing at 1200 K. $FeO_x$ films were deposited onto the Pt(111) surface at room temperature (RT) using an e-beam evaporator in a $O_2$ partial pressure of $1 \times 10^{-7}$ Torr. Subsequent annealing at higher temperatures and $O_2$ pressures led to the formation of a well-ordered $FeO_x$ film, with a coverage controlled by the deposition time.

## 3 Results and Discussion:

STM images of the $FeO_x$ monolayer, the thin films of $Fe_3O_4$ and $Fe_2O_3$ prepared as described in the experimental section are shown in **Fig. 1**. The STM images of FeO overlayer display a moiré pattern with a periodicity of approximately 2.5 nm (**Fig. 1a, inset**) originated from the lattice mismatch between the Fe-O layer (3.1 Å) and Pt substrate (2.78 Å), showing three regions of different contrasts: HCP, FCC, and TOP.[18, 19] We then prepared $Fe_3O_4$(111) thin film (~ 4.2 nm thickness) by cycles of iron deposition and oxidation at ~ 870 K in $1\times 10^{-6}$ Torr $O_2$.[40] **Fig. 1c** shows STM images of as-prepared $Fe_3O_4$(111) surface exhibits hexagonal lattice with a 5.8 Å spacing., i.e. (2× 2) structure and the surface is terminated by the tetrahedral Fe layer [40, 41]. Increasing the oxygen pressure to $5\times 10^{-5}$ Torr while annealing at the same temperature results in further oxidation of the $Fe_3O_4$(111) surface. This oxidation significantly decreases the films conductivity, hindering stable imaging at sample biased below 2.0 V (data not shown). It may suggest the formation of $Fe_2O_3$ phase as its lower conductivity compared with the $Fe_3O_4$ phase, which also agrees with the previous STM results. [20, 42] Accordingly, **Fig. 1e** displays an STM image acquired at a positive bias of 3.0 V, revealing complex moiré patterns with a periodicity of approximately 3.3 nm. High-resolution imaging resolves atomic features with an interatomic spacing of 3.0 Å (**Fig. 1e inset**). These STM characteristics suggest the formation of a bi-phase $Fe_2O_3(0001)$ surface, although its precise termination remains to be fully elucidated. [43, 44]

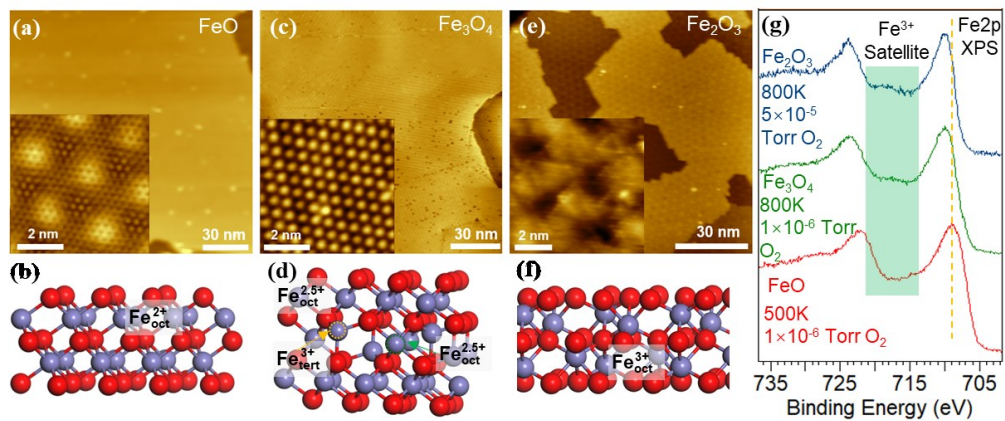

**Fig. 1 STM and XPS of the $FeO_x$/Pt(111).** Large scale STM image and structural models showing the atomic flat terrace and side view of these Fe oxide films crystal structures: (**a-b**) FeO, (**c-d**) $Fe_3O_4$(111) and (**e-f**) $Fe_2O_3$(0001) film; **Insets**: zoom-in STM image with atomic resolution. (g) showing the Fe2p XP spectra of FeO (**red curve**), $Fe_3O_4$(111) (**green curve**) and $Fe_2O_3$(0001) (**blue curve**). The orange line marking the Fe2p 3/2 XPS peak position of FeO located at 709.0 eV and shadowed green box highlighting the satellite peak region. Scanning parameters: (**a**) inset: Vs =+0.02 V, It=6.02 nA; (**c**) inset: Vs =+0.9V, It=0.7 nA; (**e**) inset: Vs =+3.08 V, It=5 nA.

XPS experiments were performed to complement the above assignment of the $FeO_x$ phase (**Fig. 1g**). The Fe2p XP spectra of ~15 ML $FeO_x$ film after annealing at ~500K under $1\times 10^{-6}$ Torr $O_2$ displays an asymmetric peak centered at 709.0 eV, characteristic of the FeO (**Fig. 1g**, **red curve**). After oxidation temperature further increased to ~ 800 K, the binding energy (BE) of this Fe2p 3/2 peak further shifted to 710.1 eV (**Fig. 1g, green curve**), in agreement with the that of $Fe_3O_4$[39]. No further peak shift was observed as the $O_2$ pressure increased to $5\times 10^{-5}$ Torr except the emergence of a shake-up satellite peak at 718.5 eV (**Fig. 1g**, **blue curve**), indicative of the formation of $Fe_2O_3$. [45] In summary, the $FeO_x$ film evolves from FeO to $Fe_3O_4$ to $Fe_2O_3$ in response to the increasing $O_2$ chemical potential.

As demonstrated above, the accurate $Fe^{2+}/Fe^{3+}$ ratio in mixed Fe oxide phases is difficult to determine due to the strong overlap of Fe 2p XP spectra among $FeO_x$ [45]. Here, we utilize Resonant Photoemission Spectroscopy (ResPES) for this purpose. Usually the enhancement in the valence band (VB) signal in ResPES originates in a two-electron process[46-52], where the Fe 2p core-level electrons are first excited to the unoccupied 3d orbitals of the conduction band (CB), then decays back to the core level, simultaneously ejecting an occupied 3d electron from the VB (**Fig. 2a**). Thus, it combines X-ray absorption process and Auger decay process together. On the practical level, partial electron yield X-ray absorption spectra (PEY-XAS) were first conducted to identify the Fe L-edge region. Then, a series of valence band spectra (VBs) as a function of photon energy within this region were captured, creating a 2D heatmap of VBs. **Fig. 2b-2c** display Fe L-edge XAS and the 2D VBs heatmap of FeO and $Fe_2O_3$ films, respectively. It clearly shows that the highest resonant enhancement in VB for $Fe^{2+}$ is at photon energy of 708.0 eV, while for $Fe^{3+}$, it is at 709.5 eV, accompanied by a secondary resonant feature at 708.0 eV. **Fig. 2d-2e** show the specific VBs of FeO and $Fe_2O_3$ at 709.5 eV ($Fe^{3+}$-resonance), 708.0 eV ($Fe^{2+}$-resonance), and 706.0 eV (off-resonance), respectively. For FeO, two resonant peaks were observed at ~4.5 eV and ~12 eV (**Fig. 2d**). For $Fe_2O_3$, three resonant peaks were observed at ~2 eV, ~4.5 eV, and ~12 eV (**Fig. 2e**). The peak at ~21eV showing in ResPES of both FeO and $Fe_2O_3$ responds to O 2s level.[53]

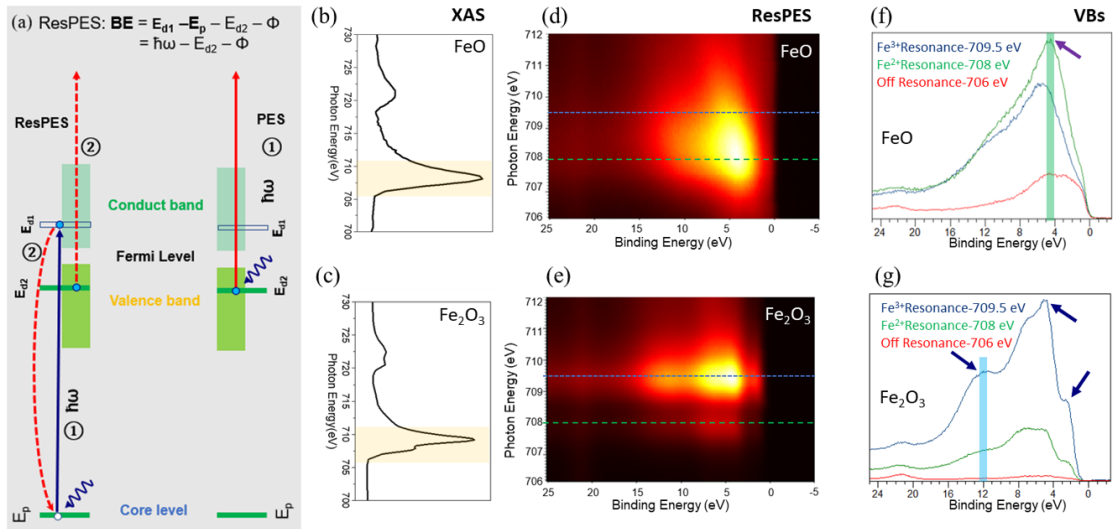

**Fig. 2 ResPES of FeO and $Fe_2O_3$ film supported on Pt(111)**. (**a**) schematic illustration showing the two-electron process of the ResPES (left) and normal valence band spectra process (right). (**b-e**) XAS and 2D heat map of VBs of FeO and $Fe_2O_3$, respectively. (**f-g**) $Fe^{3+}$-resonance (709.5 eV), $Fe^{2+}$-resonance (708.0

eV) and off-resonance VBs (706 eV) of FeO and $Fe_2O_3$, respectively. The arrows in the **(f)** and **(g)** mark the resonant peaks of $Fe^{2+}$ and $Fe^{3+}$ sites in FeO and $Fe_2O_3$, respectively. The stripes indicate the specific resonant peak positions of $Fe^{2+}$ (4.5 eV) and $Fe^{3+}$ (12 eV), which are used in the compositional analysis of the mixed $FeO_x$.

Next, we use a stoichiometric $Fe_3O_4$ film, where the $Fe^{3+}/Fe^{2+}$ concentration ratio is known as 2, to extract the linear factor (y) between resonant enhancement, $D(Fe^{n+})$, and concentration, $N(Fe^{n+})$. **Fig. 3a** displays the Fe L-edge PEY-XAS of $Fe_3O_4$, with pre-edge and main absorption peaks similar to those of $Fe_2O_3$, but with different peak ratios. The VBs heat map show similar strongest resonant features at 709.5 eV (**Fig. 3b**). Specifically, the $Fe_3O_4$ VBs features at $Fe^{3+}$-resonance (709.5 eV) and $Fe^{2+}$- resonance (708.0 eV) resemble those of $Fe_2O_3$, except for the missing of 2 eV peak at $Fe^{3+}$-resonance (**Fig. 3c**). Accordingly, we use the 12 eV peak at 709.5 eV and the 4.5 eV peak at 708.0 eV as resonant signatures of $Fe^{3+}$ and $Fe^{2+}$ cations. Notably, VBs of FeO (pure $Fe^{2+}$) in **Fig. 2d** exhibit a discernible peak intensity at 4.5 eV BE as photon energy set at 709.5 eV ($Fe^{3+}$- resonance), which is 0.68 times that of the $Fe^{2+}$ resonant peak intensity at 708.0 eV. Similarly, the VBs of $Fe_2O_3$ (pure $Fe^{3+}$) in **Fig. 2e** also display a visible shoulder at 12 eV BE as photon energy set at 708.0 eV ($Fe^{2+}$-resonance), which is 0.26 times that of the $Fe^{3+}$ resonant peak at 709.5 eV. To achieve a more precise compositional $Fe^{3+}/Fe^{2+}$ ratio, we account for these residual effects rather than relying solely on the proportional relationship between the concentration ratio $N(Fe^{3+})/N(Fe^{2+})$ and the resonant enhancement ratio (RER), $D(Fe^{3+})/D(Fe^{2+})$[48, 51]. The modified equations are listed:

$$\text{For } Fe^{3+}: D(Fe^{3+}) = y_1 * N(Fe^{3+}) + y_2 * N(Fe^{2+}) * 0.68 \quad \text{—— Equation (1)}$$

$$\text{For } Fe^{2+}: D(Fe^{2+}) = y_1 * N(Fe^{3+}) * 0.26 + y_2 * N(Fe^{2+}) \quad \text{——Equation (2)}$$

Based on the known stoichiometric of $Fe_3O_4$ and measured value of resonant enhancement value, the $y_1$ (linear factor for $Fe^{3+}$) and $y_2$ (linear factor for $Fe^{2+}$) is determined as 0.55 and 0.46. According, these equations set changes to

$$\text{For } Fe^{3+}: D(Fe^{3+}) = 0.55 * N(Fe^{3+}) + 0.31 * N(Fe^{2+}) \quad \text{—— Equation (3)}$$

$$\text{For } Fe^{2+}: D(Fe^{2+}) = 0.14 * N(Fe^{3+}) + 0.46 * N(Fe^{2+}) \quad \text{—— Equation (4)}$$

Furthermore, for other nonstoichiometric $FeO_x$ phase, the $N(Fe^{3+})/N(Fe^{2+})$ can be quantitively determined with measured $D(Fe^{3+})$ and $D(Fe^{2+})$ through solving the system of linear equations with two variables.

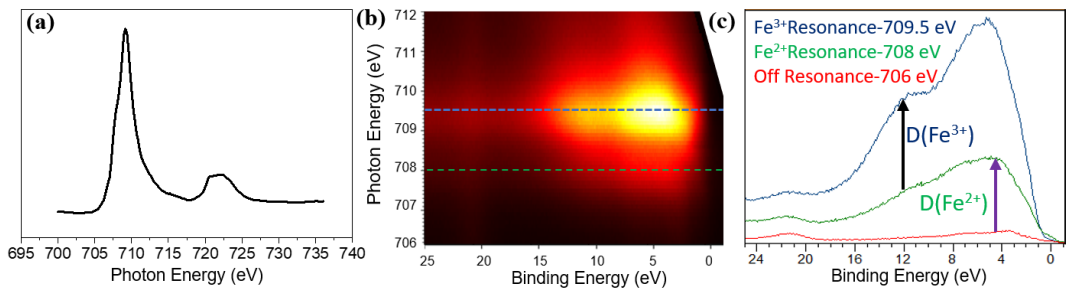

**Fig. 3 ResPES of stoichiometric $Fe_3O_4$ supported on Pt(111)**. (**a**) Fe L-edge XAS of $Fe_3O_4$. (**b**) 2D heat map of VBs of $Fe_3O_4$. (**c**) $Fe^{3+}$-resonance (709.5 eV), $Fe^{2+}$-resonance (708.0 eV) and off-resonance VBs (706 eV) of $Fe_3O_4$. Some data points near the Fermi level are absent for photon energies above 709.8 eV due to the setting of data acquisition parameters, but this does not affect the overall analysis.

Finally, we utilize ResPES to determine the valence state of the Pt supported 1ML $FeO_2$ film. The $FeO_2$ phase, first reported by the Freund group [31] and followed by other groups [26, 38, 54], forms after annealing monolayer-thick FeO at ~450K in ~15 Torr $O_2$ because of extra oxygen layer intercalation at the Pt-FeO interface. The corresponding STM image displays a higher contrast and more corrugated surface compared to FeO [26, 38, 54]. This tri-layer $FeO_2$ (O-Fe-O) is believed to be the active phase for $O_2$-rich CO oxidation via the Mars-van Krevelen (MvK) mechanism [31, 33, 34], where the intercalated oxygen layer can be reacted by CO and regenerated by $O_2$. DFT simulations suggest that the oxidation state of $FeO_2$ is approximately 3+ rather than 4+ due to back-charge transfer from the Pt substrate to the oxide layer[35, 54]. However, experimental confirmation of this charge state has predominantly relied on Fe 2p XPS, which is ambiguous for mixed $FeO_x$ phases due to overlap between $Fe^{2+}$ and $Fe^{3+}$ species[26, 39]. Moreover, the atomic structure of this tri-layer $FeO_2$ is primarily inferred from STM imaging and DFT simulations without extra experimental evidence [32]. As shown below, ResPES proves to be a powerful tool to determine the charge ratio and provide insights into the structural determination of $FeO_2$.

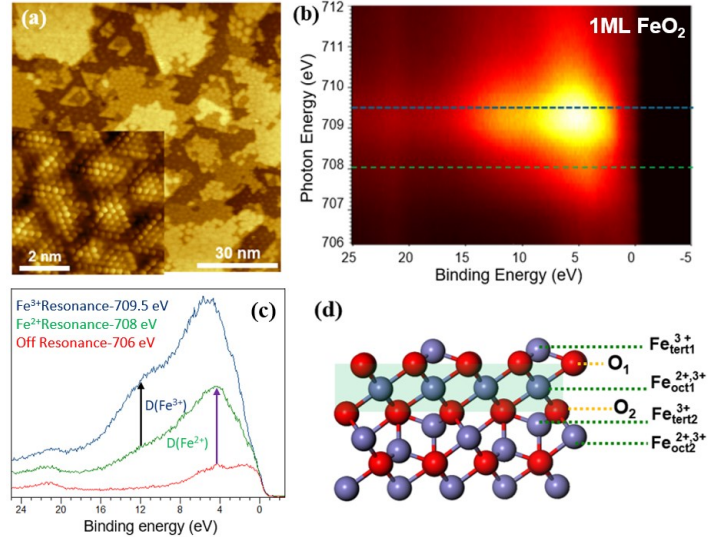

**Fig. 4 ResPES of ML FeO$_2$ film supported on Pt(111)**. (**a**) STM image of the Pt(111) supported FeO$_2$ film. Inset showing the atomic resolution STM with 3.2 Å lattice spacing. (**b**) 2D heat map of VBs of FeO$_2$ film of monolayer thickness. (**c**) Fe$^{3+}$-resonance (709.5 eV), Fe$^{2+}$-resonance (708.0 eV) and off-resonance VBs (706 eV) of FeO$_2$. (d) structural model showing the side view of Fe$_3$O$_4$(111) with Fe$_{tert1}$ layer termination. Scanning parameters: (**a**) inset: Vs =+1.0 V, It=5.0 nA.

**Fig. S1** displays the Fe 2p XPS spectra of a monolayer FeO (red curve), with a characteristic peak at 709.1 eV. After oxidation at 470K under 5×10$^{-5}$ Torr O$_2$, the Fe 2p XPS remains unchanged. Increasing the pressure O$_2$ temperature to 100 mTorr and annealing at 540 K results in FeO$_2$ formation, indicated by a blue shift of Fe 2p3/2 peak to 710.1 eV, consistent with previous literature[26, 36]. **Fig.4a** display the large scale STM image of the atomic flat FeO$_2$ film with 3.2 Å lattice spacing shown in the atomic resolution image (**inset**), agreeing with reported results[26, 32, 38]. Further Fe L-edge XAS (**Fig.S2**) and VB heat maps of this Pt (111) supported FeO$_2$ film show significant similarities to those of Fe$_3$O$_4$ phase in **Fig.3b**, suggesting that the similarity between FeO$_2$ and the Fe$_3$O$_4$ (111) (sub)lattice. However, the corresponding Fe$^{3+}$-ResPES displays a weaker peak intensity at 12 eV, indicating a lower Fe$^{3+}$/ Fe$^{2+}$ratio compared to Fe$_3$O$_4$. Based on the Equations (3)-(4), we determine the N(Fe$^{3+}$)/N(Fe$^{2+}$) ratio of FeO$_2$ film to be unit. This equal composition ratio of Fe$^{3+}$ and Fe$^{2+}$ species is also valid among the sub-monolayer FeO$_2$ film (**Fig. S3**).

To further elucidate the structural relationship between FeO$_2$ and Fe$_3$O$_4$, the **Fig. 4d** shows the side view of Fe$_3$O$_4$(111) terminated structures with stacking along the [111] direction. From top to bottom, its unit cell consists of a tetrahedral Fe layer (Fe-tet1, Fe$^{3+}$), oxygen layer (O$_1$), octahedral Fe layer (Fe-oct1, 50% Fe$^{3+}$ + 50% Fe$^{2+}$), oxygen layer (O$_2$), another tetrahedral Fe layer (Fe-tet 2, Fe$^{3+}$), and another octahedral Fe layer (Fe-oct 2, 50% Fe$^{3+}$ + 50% Fe$^{2+}$).[40] Given that previous theoretical calculations proposed an O-Fe-O stacking sequence for the FeO$_2$ structure [32, 35] and considering the 1:1 ratio of Fe$^{3+}$/Fe$^{2+}$ experimentally resolved in RPES measurement, we propose that Pt(111)-supported FeO$_2$ structure is analogous to the O$_1$-Feoct1-O$_2$ sublattice within Fe$_3$O$_4$(111) shown in **Fig. 4d**, where octahedral Fe cations are coordinated by upper (O$_1$) and lower (O$_2$) oxygen layers. This proposed FeO$_2$ structure (O$_1$-Feoct1-O$_2$ interfacial configuration) is in good agreement with the previous DFT simulations.

**Conclusion**

In summary, series of Fe oxides on Pt(111) were prepared after post annealing at different oxygen atmosphere and temperature, which severs as the standard stoichiometric oxides, i.e., FeO, Fe$_2$O$_3$ and Fe$_3$O$_4$, to establish the ResPES methodology for nonstoichiometric FeO$_x$. Utilizing ResPES on catalytically active FeO$_2$ phase, we demonstrated that this FeO$_2$ is composed of equal amount of Fe$^{3+}$ and Fe$^{2+}$ species with an average +2.5 valence. Accordingly, this FeO$_2$ phase may evolve from the Fe$_3$O$_4$ (111) sublattice, consisting of the octahedral Fe1 layer bonded to upper (O1) and lower oxygen (O$_2$) layers. This powerful ResPES method demonstrated here can be applied to practical Fe oxide nanoparticle systems to in-situ determine composition ratio of Fe$^{3+}$/Fe$^{2+}$ with high precision.

**Acknowledgement**

This work is supported by the Office of Basic Energy Sciences (BES; Chemical Sciences, Geosciences, and Biosciences Division) of the U.S. Department of Energy (DOE) under Contract DE-AC02-05CH11231, FWP CH030201 (Catalysis Research Program). It is also financially supported by National Natural Science Foundation of China (22376194), National Key Research and Development Program (2022YFA1505500), Natural Science Foundation of Fujian Province,

**Contributions**

M.S. oversaw the project. H.C., conduct the X-ray spectroscopy experiment supported by the beamline scientists (S.N.). Y.L., X. Zhang, and S. Zhao., did the STM experiment. H.S.L D wrote the python code for the reading of 2D ResPES data. Data analysis were done by H.C., S.N. and M.S. The paper was written by H.C. with contributions from all authors. Y.L., S.N. and M.S. helped to review and edit the manuscript.